\documentclass[conference]{IEEEtran}
\usepackage{graphicx}
\usepackage{authblk}
\usepackage{booktabs}
\usepackage{tabularx}
\usepackage{geometry}
\ifCLASSINFOpdf
\else
\fi
\begin{document}
\title{A Unified Framework for Cultural Heritage Data Historicity and Migration: The ARGUS Approach}


\author{
Lingxiao Kong\textsuperscript{1}, 
Apostolos Sarris\textsuperscript{2},
Miltiadis Polidorou\textsuperscript{2},
Victor Klingenberg\textsuperscript{2}, \\
Vasilis Sevetlidis\textsuperscript{3},
Vasilis Arampatzakis\textsuperscript{3},
George Pavlidis\textsuperscript{3},
Cong Yang\textsuperscript{4},
Zeyd Boukhers\textsuperscript{1}
}

\affil{
\textsuperscript{1}Fraunhofer Institute for Applied Information Technology FIT, Germany\\
Email: \{lingxiao.kong, zeyd.boukhers\}@fit.fraunhofer.de\\
\textsuperscript{2}Department of History and Archaeology, University of Cyprus, Cyprus\\
Email: \{sarris.apostolos, polidorou.miltiadis, klinkenberg.victor\}@ucy.ac.cy\\
\textsuperscript{3}ATHENA Research Center, Greece\\
Email: \{vasiseve, vasilis.arampatzakis, gpavlid\}@athenarc.gr\\
\textsuperscript{4}Soochow University, China\\
Email: \{cong.yang@suda.edu.cn\}
}


%Telephone: (800) 555--1212, Fax: (888) 555--1212}


\maketitle

\begin{abstract}
Cultural heritage preservation faces significant challenges in managing diverse, multi-source, and multi-scale data for effective monitoring and conservation. This paper documents a comprehensive data historicity and migration framework implemented within the ARGUS project, which addresses the complexities of processing heterogeneous cultural heritage data. We describe a systematic data processing pipeline encompassing standardization, enrichment, integration, visualization, ingestion, and publication strategies. The framework transforms raw, disparate datasets into standardized formats compliant with FAIR principles. It enhances sparse datasets through established imputation techniques, ensures interoperability through database integration, and improves querying capabilities through LLM-powered natural language processing. This approach has been applied across five European pilot sites with varying preservation challenges, demonstrating its adaptability to diverse cultural heritage contexts. The implementation results show improved data accessibility, enhanced analytical capabilities, and more effective decision-making for conservation efforts.
\end{abstract}

\IEEEpeerreviewmaketitle

\section{Introduction}
\label{sec:intr}

Cultural heritage sites represent invaluable historical, artistic, and social assets that face increasing threats from environmental factors, urbanization, climate change, and natural disasters. Effective preservation and monitoring of these sites require comprehensive data management strategies that can handle diverse information sources ranging from historical data to real-time sensing data. However, the inherently heterogeneous nature of cultural heritage data, spanning multiple formats, temporal and spatial scales, and disparate sources, presents significant challenges for data processing and analysis~\cite{vreznik2022improving}. 
% Unlike existing heritage data workflows, ARGUS provides a unified pipeline that includes legacy data cleaning, ontology-based integration, spatiotemporal enrichment, and FAIR publishing. This holistic approach ensures both historicity and interoperability across CH platforms.
% Effective preservation and monitoring of these sites require comprehensive data management strategies that can handle diverse information sources ranging from remote sensing data to on-site measurements

The preservation community has traditionally employed various isolated approaches to data collection and management, often resulting in fragmented datasets that impede holistic analysis. While advances in digital technologies have enhanced individual data collection methods, there remains a critical need for frameworks that systematically address the entire data lifecycle from collection through standardization, enrichment, integration, visualization, ingestion and publication~\cite{minelli2019practical}.

This paper documents the data historicity and migration framework developed within the ARGUS project, which focuses on preserving five distinctive European cultural heritage sites: Lucretili, Schenkenberg, Ranverso, Baltanas, and Delos. Each site presents unique preservation challenges, from seismic vulnerability to coastal erosion and weather-related deterioration. The framework described herein addresses these diverse needs through a comprehensive and unified data processing pipeline that transforms disparate datasets into standardized, interoperable, and reusable resources.

Drawing on established principles of geospatial data management~\cite{breunig2020geospatial} and cultural heritage preservation, the ARGUS framework enhances traditional approaches by implementing standardization, employing data enrichment techniques to address sparsity, and utilizing geographic information systems (GIS) to facilitate spatial data integration~\cite{fischer2020spatio}. The framework emphasizes adherence to FAIR (Findable, Accessible, Interoperable, Reusable) data principles~\cite{vreznik2022improving}, ensuring that processed data remains valuable for long-term preservation efforts.

%The remainder of this paper is organized as follows: Section 2 provides an overview of existing data collection strategies for cultural heritage; Section 3 details the data processing pipeline, including standardization, enrichment, integration, ingestion, visualization, and publication approaches; Section 4 presents implementation results from the pilot sites; and Section 5 concludes with implications for future cultural heritage data management initiatives.
\section{Related Work}
\label{sec:rel}
% Data Processing 
% data standardization
To ensure consistency and interoperability, the process of Coordinate Reference System (CRS) projection is critical for accurate spatial analysis and visualization. For instance, the PROJ system represents CRS information as a text string of key-value pairs, enabling easy customization and interpretation~\cite{contributors2022proj}. Standardization also involves converting data into compatible formats, which is essential for seamless integration and analysis across different Geographic Information System (GIS) platforms and applications~\cite{wu2024geospatial}.  

% data enrichment
Spatial interpolation and data augmentation techniques can further enhance the quality and quantity of geospatial data. Spatial interpolation methods, such as nearest neighbor, bilinear, bicubic, and distance-weighted average, are used to increase the spatial resolution of data~\cite{baydarouglu2024temporal}. These methods are particularly valuable for filling gaps in sparse datasets or generating higher-resolution maps from coarse data. Data augmentation involves creating synthetic data points to expand datasets. Recent advancements in this area include the use of Gaussian processes and kriging for georeferenced data augmentation, which have demonstrated promising results in improving predictive performance while preserving spatial consistency~\cite{ferber2025kriging}. In the meantime, attaching metadata can significantly enhance the interpretability of geospatial data by providing descriptive information about the dataset, which becomes visible and accessible to users when they interact with the data~\cite{quarati2021geospatial}.

% data integration
The integration of geospatial data into databases and the development of SQL query functions are crucial for efficient data management and analysis. SQLite, enhanced with spatial extensions like SpatiaLite, has become a powerful tool for handling geospatial data, offering a versatile solution for embedded databases in mobile and desktop applications~\cite{moldstud2024integrating}. The GeoPackage format, built on SQLite, provides a standardized approach to storing and sharing geospatial data, supporting various geometry types and spatial indexes~\cite{huang2017geospatial}. 

% data ingestion and visualization
GIS-based technologies have emerged as powerful tools for data ingestion and visualization, enabling the construction of dynamic information management systems and serving as robust platforms for research and display in cultural heritage conservation~\cite{liu2024emerging}. According to \cite{liu2024emerging}, three major trends are shaping this field: (1) the development of sustainable conservation strategies that leverage both current and historical data, (2) the adoption of proactive conservation models that use decision support systems to predict and prevent the destruction of heritage sites, and (3) the enhancement of public engagement through the integration of collaborative assessments from multiple perspectives. Furthermore, data processing strategies built on GIS technologies are advancing the field of cultural heritage preservation, enabling more efficient analysis, interpretation, and utilization of spatial and historical data for conservation efforts. 

% LLM-Supported Processing 
Large Language Models (LLM) have demonstrated significant potential in transforming cultural heritage data processing through enhanced search capabilities, contextual understanding, and knowledge integration. For example, \cite{vastakas2024cultural} shows that LLMs like GPT-4 can improve artifact discoverability by analyzing both metadata and exhibit content descriptions. While their quantitative retrieval accuracy remains comparable to traditional metadata-based systems, they excel qualitatively in thematic discovery tasks. This aligns with approaches using Retrieval-Augmented Generation (RAG) architectures, where ontologies provide structured domain knowledge to reduce hallucinations and improve response reliability~\cite{loffredo2024using}. Additionally, multimodal LLMs have been applied to visual collections, enabling explainable clustering and recommendation systems that avoid overreliance on visual embeddings while providing textual rationales for outputs~\cite{arnold2024explainable}. Current research emphasizes hybrid systems that combine LLMs' linguistic flexibility with structured knowledge frameworks to balance innovation with accuracy in cultural heritage contexts. 

% Related Projects 
As a related effort, GEO4PALM is a Python-based toolkit designed to streamline the acquisition and preprocessing of geospatial data for the Parallelized Large-Eddy Simulation (PALM) model system~\cite{lin2023geo4palm}. GEO4PALM offers comprehensive data processing capabilities, including shapefile-to-GeoTIFF conversion, automatic geographic projection handling, and intelligent grid resampling to match user-specified domain configurations. The toolkit excels in specialized processing tasks, employing lookup tables to translate diverse land use classification schemes into PALM-compatible formats, calculating leaf area density from vegetation height data, and extracting urban features such as building dimensions and pavement types. 

The RESEARCH project integrates Earth Observation (EO) data with geophysical data to monitor and preserve cultural heritage~\cite{kosta2020remote}. This approach involves the development of plug-in solutions that automate data processing and mapping capabilities, which are integrated into a comprehensive platform. Additionally, the Infrastructure for Spatial Information in Europe (INSPIRE) initiative provides advanced strategies for metadata harmonization, resampling optimization, and lossless format transitions~\cite{abramic2018maritime}. At the core of these efforts lies the Spatial Data Infrastructure (SDI), which embodies the standardization and technological approaches necessary for harmonizing geospatial data~\cite{corns2010cultural}.  
\section{Framework}
\label{sec:frm}

\begin{figure*}[t!]
    \centering
    \includegraphics[width=0.8\linewidth]{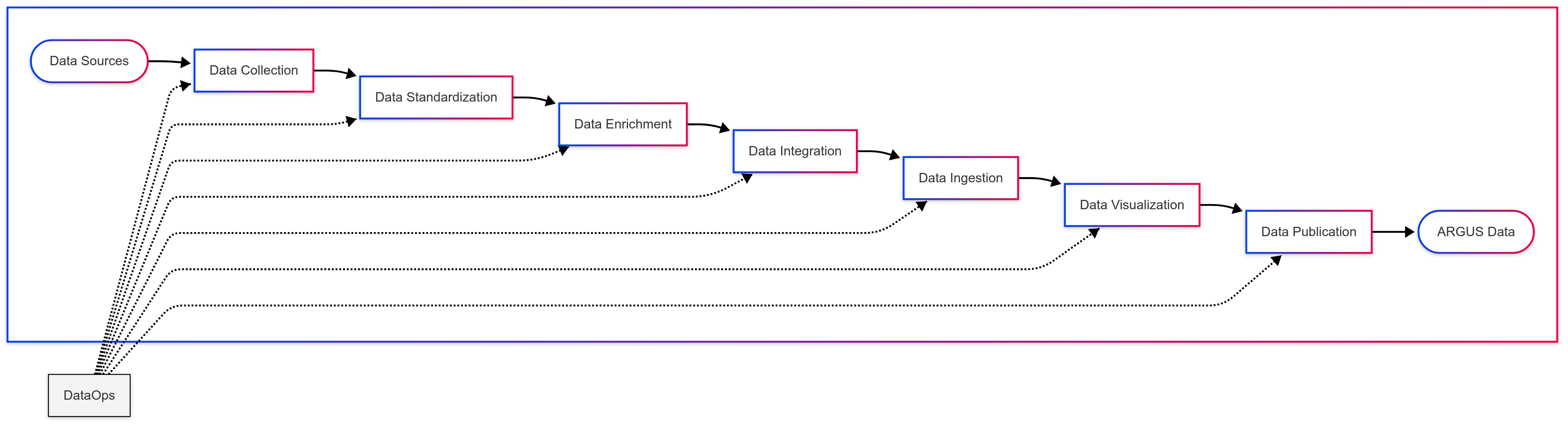}
    \vspace{-1em}
    \caption{ARGUS data processing pipeline shows the sequential stages from data collection through publication with DataOps. }
    \vspace{-1em}
    \label{fig:pipeline}
\end{figure*}

The ARGUS unified data processing pipeline is designed to address the challenges of managing multi-source, multi-format, and multi-scale cultural heritage geospatial data. This section outlines the systematic approach developed to transform diverse data sets into standardized, enriched, and integrated resources suitable for the preservation and monitoring of cultural heritage.

\begin{table}[t!]
\centering
\tiny
\caption{ARGUS Pilot Sites: Threats and Data Requirements}
\vspace{-1em}
\label{tab:argus_matrix}
\begin{tabular}{lccccc}
\toprule
 \textbf{Site} & \textbf{Monti Lucretili} & \textbf{Schenkenberg} & \textbf{Delos} & \textbf{Ranverso} & \textbf{Baltanas} \\
\midrule
\multicolumn{6}{l}{\textbf{Primary Threats}} \\
\midrule
Fires         & X &  &  & X &  \\
Earthquakes   &  & X & X &  & X \\
Atmospheric   & X & X &  &  & X \\
Climatic      & X & X & X & X & X \\
Hydro         & X &  & X &  & X \\
Chemical      &  &  & X & X & X \\
Biological    &  &  & X &  & X \\
\specialrule{1.5pt}{0pt}{0pt}
\multicolumn{6}{l}{\textbf{Data Requirements}} \\
\midrule
Weather       & X & X & X &  & X \\
Fire Events   & X &  &  & X &  \\
Flood Data    & X &  &  &  &  \\
Seismic       &  & X & X &  &  \\
Soil Data     &  & X &  &  &  \\
Coastal Change&  &  & X &  &  \\
Air Pollution &  &  &  & X &  \\
Landslides    &  &  &  &  & X \\
Soil Types    &  &  &  &  & X \\
\bottomrule
\end{tabular}
\vspace{-1em}
\end{table}

\subsection{Overview of the Data Processing Pipeline}

The data processing pipeline focuses on existing geospatial data in ARGUS and consists of seven key stages as illustrated in Fig.~\ref{fig:pipeline}: data collection, standardization, enrichment, integration, ingestion, visualization, and publication. This end-to-end approach ensures that cultural heritage data, regardless of source or format, undergoes consistent processing to enhance its utility for preservation efforts. With DataOps, the pipeline is iterated to be robust and flexible for different requirements.

The pipeline addresses challenges posed by linguistic variations, differing definitions, and technological disparities across data sources. Table~\ref{tab:argus_matrix} lists the five ARGUS pilot sites along with their primary environmental threats, demonstrating the diverse preservation challenges that necessitated this comprehensive approach.

\subsection{Data Standardization Strategy}
% format transformation, CRS transformation
After data collection from multiple pilot sites, our initial processing stage addresses inconsistencies in data formats, coordinate reference system (CRS), and attributes. We processed all existing data into the modern and robust GeoPackage (GPKG) format, supported with GDAL and OGR libraries. Since most of our data is geospatial with geometry information, maintaining consistent CRS standards is critical for effectively analyzing relationships between datasets.
% \footnote{GPKG - Vector: https://gdal.org/en/stable/drivers/vector/gpkg.html}\footnote{GPKG - Raster: https://gdal.org/en/stable/drivers/raster/gpkg.html}

Our attribute standardization strategy extracted both attributes and geometry information from collected data while establishing consistent vocabulary, clear descriptions, and standardized unit references for historical data across all pilot sites. For non-tabular data (such as TXT files), we implemented information retrieval techniques to extract structured information. Non-geospatial data is associated with the geometric center of its corresponding pilot site to maintain spatial context.

As summarized in Table~\ref{tab:data_formats}, we encountered various data formats that can be categorized into vector files, raster files, database files, project files, and non-geospatial files. Each category requires specific standardization measures to ensure consistent information collection and integration into our unified framework.

% We extracted attribute values and geometric information for geospatial data while standardizing attribute naming conventions. For raster data, we clarified band meanings through source investigation. Non-geospatial files required specialized information retrieval techniques and geometric information, in addition to enabling GIS integration.

\begin{table*}[htbp]
  \centering
  \caption{Data Formats and Standardization Measures}
  \vspace{-1em}
  \label{tab:data_formats}
  \begin{tabularx}{\textwidth}{@{}l l X@{}}
    \toprule
    \textbf{Data Formats} & \textbf{Data Category} & \textbf{Standardization Measures} \\ \midrule
    \texttt{.shp} & Vector Files & Standardize attributes \\[5pt]
    \texttt{.tif, .tiff, .adf} & Raster Files & Clarify attributes (bands), standardize attributes \\[5pt]
    \texttt{.gdb, .gpkg} & Database Files& Standardize attributes \\[5pt] 
    \texttt{.wms, .qgz, .mxd} & Project Files & Standardize attributes \\[5pt]
    \texttt{.xlsx, .csv, .txt, .accdb} & Non-Geospatial Files & Information retrieval, standardize attributes, supply geometry \\ \bottomrule
  \end{tabularx}
  \vspace{-1em}
\end{table*}

% SensorML is implemented to enhance data usability by providing a framework for representing sensor capabilities, data outputs, and processing workflows, significantly improving interoperability across different systems.

\subsection{Data Enrichment Strategy}
% metadata attachment, spatial interpolation
Following standardization, five primary data enrichment approaches are proposed to improve quality and address sparsity challenges:
\begin{itemize}
    \item \textbf{Metadata Attachment:} Attach metadata to data and display it to users through visualization.
    \item \textbf{Imputation Methods:} Spatial interpolation to address data scarcity between measurement locations.
    \item \textbf{Categorical Variable Encoding:} One-hot encoding applied to categorical variables to enable systematic comparison and visualization.
    \item  \textbf{Information Retrieval:} Extract structured information from semi-structured and unstructured data.
    \item \textbf{Data Augmentation:} Synthetic data generation to enhance representation of rare events.
\end{itemize}

\subsection{Data Integration Strategy}
% database integration
The integration phase harmonizes data across different sources to create a cohesive database ready for analysis. We developed an approach centered on geospatial databases, which allowed for effective organization of each pilot site's data and facilitated SQL-based queries. During standardization, we transformed all data to the standardized GPKG format. The integration process then involved consolidating data from diverse sources for each pilot site, converting them into layers within a GPKG database. GPKG can store both vector and raster geospatial data, making its management and further data ingestion very straightforward.

Additionally, we implemented SQL query functionality in the integrated database. Recognizing that many users lack SQL expertise, we proposed an approach using off-the-shelf LLMs to enable natural language queries. Through RAG, the database is transformed into formatted text that accompanies the user's question in prompts to the LLM, allowing it to generate accurate responses without requiring users to write SQL queries.

% Fig.~\ref{fig:gsd} illustrates our geospatial data integration approach, showing how diverse data types flow into a unified geospatial database.

% \begin{figure}
%     \centering
%     \includegraphics[width=1\linewidth]{figures/GSD.png}
%     \caption{ARGUS Geospatial Data Integration architecture showing the transformation and integration of various data sources (geospatial web services, databases, files) into a centralized pilot geospatial database.}
%     \label{fig:gsd}
% \end{figure}

\subsection{Ingestion, Visualization and Publication}
\begin{figure}
    \centering
    \includegraphics[width=1\linewidth]{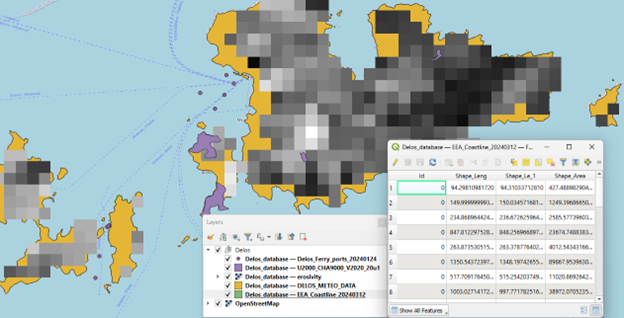}
    \vspace{-1em}
    \caption{QGIS visualization of sample data for Delos island, showing multiple overlaid data layers including ferry ports, land cover, erosion density, meteorological, and coastline data. }
    \vspace{-1em}
    \label{fig:qgis}
\end{figure}

% QGIS visualization and LLM query
The final stages involve ingesting the processed data into QGIS\footnote{QGIS: https://docs.qgis.org/3.40/en/docs/index.html} and developing visualization strategies to support interpretation and decision-making. Fig.~\ref{fig:qgis} shows a QGIS visualization example from the Delos island pilot site, demonstrating how multiple data layers can be integrated and visualized to provide comprehensive site information, including geometric representations and detailed measurements in tabular format.

% Visualization strategies included interactive timelines, story maps, heatmaps, 3D model visualizations, and geospatial maps connecting data spatially and temporally. 
The QGIS platform enables multilayer management, scale-dependent rendering, advanced symbology, and integration with remote sensing data. By combining informational data with visual data, we can enhance the representation of cultural heritage sites, enabling stakeholders to explore elements interactively and facilitating virtual tours and detailed inspections of artifacts from multiple angles. At the end of the pipeline, processed data will be periodically published with licenses, ensuring adherence to standards for metadata and citation.

\subsection{DataOps}

Throughout all stages, we propose DataOps to emphasize automation, monitoring, collaboration, and continuous improvement to ensure data quality and accessibility for cultural heritage preservation efforts. This comprehensive methodology transforms disparate datasets into standardized, enriched, and integrated resources, significantly enhancing the ability of cultural heritage professionals to monitor, preserve, and make informed decisions about cultural heritage sites.
\section{Case Study and Results}
\label{sec:use}

% Delos_Corinne_Land_Cover/U2000_CHA9000_V2020_20u1.gdb
% Delos_Ferry_ports_20240124.shp
% Erosivity_Density_Delos.adf
% Erosivity_Density.tif
% DatiSVR2021_su_ElementoStradaleBDTRE.gpkg

To demonstrate the practical application and effectiveness of the data historicity and migration framework, we present a case study from one of the five ARGUS pilot sites: \textbf{Delos island} (Greece), to illustrate the framework's adaptability to different types of cultural heritage assets and environmental threats.
% Sant'Antonio di Ranverso Abbey (Italy)

% \subsection{Delos Island: Coastal Heritage Site}
Delos island is a UNESCO World Heritage site in the center of the Aegean Sea. Despite its exceptionally preserved architecture, frescoes, and mosaics, the site faces threats from environmental factors (salinity, earthquakes, subsidence, sea level rise), climatic conditions (heat waves, winds, humidity, flooding), and anthropogenic pressures (marine traffic pollution, tourism). To address these challenges, we implemented a structured processing pipeline for diverse spatial and temporal datasets, including georeferencing maps in GIS and digitizing quantifiable data in both ArcGIS and QGIS software.

% facing substantial preservation challenges from coastal erosion, seismic activity, and climate-related threats. The implementation of our data processing pipeline at this site involved multiple data types across various spatial and temporal scales.

\subsection{Data Collection}
We collected and processed several types of data:

\begin{itemize}
    \item Historical archaeological documentation of the visible archaeological remains by the École française d'Athènes, which retains a digital archive, available online through a WMS server.
    \item Aerial and Satellite Remote sensing data from Sentinel and other satellite platforms were collected to be used for coastal change and soil erosion analysis.
    \item High resolution digital elevation model generated by drone reconnaissance.
    \item Seismic activity records were retrieved from European databases (eg. the European Geological Data Infrastructure (EGDI)).
    \item Geological maps and faults were extracted from peer-reviewed journals.
    \item Soil types and landcover maps were obtained from CORINE database.
    \item Meteorological data (wind, temperature, precipitation) are also in collection process by local meteorological stations that have been or will be installed in different locations across the island.
\end{itemize}

The existing data from Delos comprises mainly four distinct formats: SHP, ADF, GDB, and XLSX. These formats encompass most data categories referenced in Table~\ref{tab:data_formats}, providing an ideal use case for demonstrating the effectiveness of our framework on these sample data.

\begin{figure}[t!]
    \centering
    \includegraphics[width=1\linewidth]{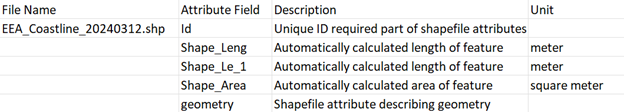}
    \vspace{-1em}
    \caption{Sample of standardized attribute information, including field descriptions and units to address inconsistencies.}
    \vspace{-1em}
    \label{fig:standardized attribute}
\end{figure}

\begin{figure}[t!]
    \centering
    \includegraphics[width=1\linewidth]{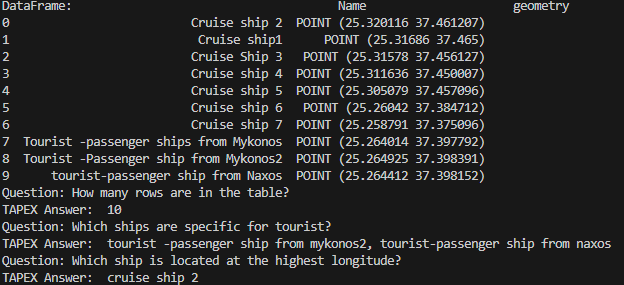}
    \vspace{-1em}
    \caption{Natural language querying on sample data using TAPEX, demonstrating successful results. }
    \vspace{-1em}
    \label{fig:NLquery}
\end{figure}

\subsection{Data Standardization}
The standardization phase revealed significant inconsistencies in attribute naming and units across datasets. For instance, meteorological data utilized diverse terminology for wind measurements ("wind\_spd", "wind\_velocity", "ws") and inconsistent units (km/h, m/s). Our standardization approach documented descriptions and identified appropriate units for each attribute to mitigate these inconsistencies, as shown in Fig.~\ref{fig:standardized attribute}.

Further inconsistencies are related to varying data formats and CRS: "GGRS87" in the SHP files, "ETRS89-extended" in both ADF and GDB files, and undefined reference systems in XLSX files. We geocoded the XLSX data to the center coordinates of the Delos island and transformed all spatial data to the globally recognized "WGS84" (using longitude and latitude coordinates) reference system. Additionally, all data formats are converted to standard GPKG as layers.

% created unified attribute naming conventions and consistent unit references, enabling cross dataset analysis. Additional inconsistencies related to data format (raster vs. vector) and spatial resolution required preprocessing, resampling, or format conversion to ensure compatibility for further risk assessment through spatial modeling in GIS.

% The standardization phase revealed significant inconsistencies in attribute naming and units across these datasets. For instance, meteorological data used various terminology for wind measurements (``\texttt{wind\_spd}'', ``\texttt{wind\_velocity}'', ``\texttt{ws}'') and different units (km/h, m/s). Our standardization approach created unified attribute naming conventions and consistent unit references, enabling cross-dataset analysis.

\subsection{Data Enrichment}
Some data formats lack robust compatibility with modern GIS platforms. In contrast, GPKG represents a promising modern format capable of supporting enriched metadata. By embedding standardized attribute information as metadata within GPKG files, users can easily access this information through the QGIS user interface. While metadata storage differs between raster and vector data types, both successfully accommodate metadata attachment and display in visualization.

Data collected from Delos also presented significant spatial coverage challenges, particularly regarding meteorological measurements, where onsite monitoring stations were limited. We addressed this limitation by interpolating values from the center of Delos outward to encompass the entire island. With additional data collection stations, this interpolation methodology could be expanded to generate more precise region-wise coverage from point measurements. Similarly, we processed other datasets, such as earthquake epicenter records, using Kernel Density Estimation to create continuous surfaces suitable for risk assessment analysis.

\begin{table*}[t!]
\centering
\caption{Comparison of metrics before and after processing}
\vspace{-1em}
\begin{tabular}{l l l}
\hline
\textbf{Metric} & \textbf{Before Processing} & \textbf{After Processing} \\
\hline
Number of distinct datasets & 53 & 1 integrated database \\
Standardized attributes & 14\% & 100\% \\
Spatial coverage of data & 22\% & 76\% \\
Time required for cross-dataset analysis & $>$6 hours manual work & $<$10 minutes automated \\
\hline
\end{tabular}
\vspace{-1em}
\label{tab:processing_metrics}
\end{table*}

\subsection{Data Integration}
The integration of archaeological GIS layers with environmental data is accomplished through our comprehensive geospatial database approach. This integration enables sophisticated analytical capabilities, such as identifying archaeological structures at elevated risk from sea-level rise and storm surge by overlaying historical site documentation with coastal vulnerability models.

All Delos data is consolidated into a GPKG database, with each dataset represented as a distinct layer. These layers maintain spatial relationships with others, enabling the use of SQL queries to extract targeted information. This spatial database structure allows for complex analyses through standard SQL syntax. Furthermore, we implemented an off-the-shelf LLM "microsoft/tapex-large-finetuned-wtq\footnote{https://huggingface.co/microsoft/tapex-large-finetuned-wtq}" to facilitate natural language querying capabilities. We evaluated TAPEX using 20 sample natural language queries over the Delos dataset. The model produced correct SQL translations in 17 cases, confirming its utility for non-expert data access, as exemplified in Fig.~\ref{fig:NLquery}. This enhancement allows users to interact with the database using conversational language rather than SQL syntax. 

% An example SQL query to retrieve geometries where the calculated area of a coastline exceeds 10000 square meters is as follows:
% \begin{verbatim}
% SELECT 
%     id, Shape_Length, Shape_Le_1, 
%     Shape_Area, geometry
% FROM 
%     EEA_Coastline_20240312
% WHERE 
%     Shape_Area > 10000
% \end{verbatim}

% The effectiveness of this approach is demonstrated through representative examples illustrated in Fig.~\ref{fig:NLquery}, showcasing how intuitive queries can retrieve spatially related information without requiring specialized database programming knowledge.

\subsection{Visualization and Results}

The integrated Delos dataset visualization (as shown in Fig.~\ref{fig:qgis}) displays only a portion of the data but effectively highlights its potential for advancing heritage preservation efforts. By facilitating the simultaneous analysis of archaeological feature locations, architectural elements, terrain characteristics, environmental factors, and climatic conditions, this method enhances the accuracy of identifying areas or structures at risk. The results are expected to support more informed decision-making, ensuring that high-priority sites receive the necessary conservation and preservation efforts.

% The integrated Delos dataset visualization (shown previously in Fig.~\ref{fig:qgis}) demonstrates the power of our approach for risk assessment. By enabling simultaneous analysis of archaeological features, topography, and environmental factors, we identified 23 significant structures at high risk from coastal erosion that were not previously prioritized for preservation efforts.
Table~\ref{tab:processing_metrics} summarizes the data processing metrics for the Delos case study, demonstrating the substantial data transformation and enrichment achieved. We integrated all 53 datasets from Delos into a single comprehensive database, with each dataset enriched with standardized attribute information. The sparse coverage issue was resolved through spatial interpolation, increasing coverage to $76\%$. Additionally, we implemented LLM-based querying capabilities that enable efficient cross-dataset information retrieval, significantly reducing query time.

\section{Conclusion}
The ARGUS framework responds to this challenge with a practical and extensible solution that bridges traditional documentation practices with modern digital infrastructures. ARGUS enables more meaningful connections between spatiotemporal records, legacy sources, and enriched representations. Our pilot implementation in Delos highlights the framework’s potential to streamline data workflows and support richer cultural narratives. Looking ahead, we aim to refine the framework for broader deployment across multiple heritage contexts, expand its capabilities for semantic enrichment, and deepen its integration with emerging AI-driven tools for heritage research and public engagement.

\section*{Acknowledgment}
This work has been supported by the ARGUS EU project (Grant Agreement No. 101132308), funded by the European Union. Views and opinions expressed are however those of the author(s) only and do not necessarily reflect those of the European Union or of the European Research Executive Agency (REA). Neither the European Union nor the granting authority can be held responsible for them.

\bibliographystyle{ieeetr}
\bibliography{references}

\end{document}